\documentclass[aps,prb,twocolumn,groupedaddress]{revtex4}
\usepackage{graphicx}
\usepackage{amsmath}

\begin{document}

% Use the \preprint command to place your local institutional report
% number in the upper righthand corner of the title page in preprint mode.
% Multiple \preprint commands are allowed.
% Use the 'preprintnumbers' class option to override journal defaults
% to display numbers if necessary
%\preprint{}

%Title of paper
\title{Hysteresis effects in the phase diagram of
multiferroic GdMnO$_3$}
% studied by thermal expansion and
% magnetostriction}

 \author{J.\ Baier,$^1$ D.\ Meier,$^1$ K.\ Berggold,$^1$ J.\ Hemberger,$^2$
 A.\ Balbashov,$^3$ J.\, A.\ Mydosh,$^1$ and T.\ Lorenz$^1$}
\email{lorenz@ph2.uni-koeln.de}
\affiliation{$^1$II.\,Physikalisches Institut, University of
 Cologne, Germany \\
 $^2$\,Institut f\"{u}r Physik, University of Augsburg, Germany \\
 $^3$\,Moscow Power Engineering Institute, Moscow, Russia}

\date{\today}

\begin{abstract}
We present high-resolution thermal expansion $\alpha(T)$ and
magnetostriction $\Delta L(H)/L$ measurements of GdMnO$_3$, which
develops an incommensurate antiferromagnetic order (ICAFM) below
$T_{\rm N}\simeq$ 42\,K and transforms into a canted A-type
antiferromagnet (cAFM) below $T_{\rm c}\simeq 20\,$K. In
addition, a ferroelectric polarization ${\bf P}||a$ is observed
below $T_{\rm FE} $ for finite magnetic fields applied along the
$b$ direction. In zero magnetic field we find a strongly
anisotropic thermal expansion with certain, rather broad
anomalous features. In finite magnetic fields, however, very
strong anomalies arise at $T_{\rm c}$ for fields applied along
each of the orthorhombic axes and at $T_{\rm FE}$ for fields
along the $b$ axis. Both phase transitions are of first-order
type and strongly hysteretic. We observe a down-bending of the
ICAFM-to-cAFM phase boundary $T_{\rm c}(H)$ for low magnetic
fields and our data give evidence for coexisting phases in the
low-field low-temperature range.

\end{abstract}

% insert suggested PACS numbers in braces on next line
\pacs{75.47.Lx,64.70.Rh,65.40.De,75.80.+q}

% insert suggested keywords - APS authors don't need to do this
%\keywords{}

%\maketitle must follow title, authors, abstract, \pacs, and \keywords
\maketitle

%\section{Introduction}
The recent discovery of very large magnetoelectric effects in the
rare-earth manganites \textsl{R}MnO$_3$ has reopened the field of
the so-called multiferroic materials.\cite{kimura03a,goto04a}
Multiferroic means, that several ferro-type orders like
ferromagnetism, ferroelectricity or ferroelastivity coexist. The
rare-earth manganites \textsl{R}MnO$_3$ may be grouped into the
hexagonal ones (\textsl{R}= Ho,$\dots$, Lu) and those with
orthorhombically distorted perovskite structures (\textsl{R}=
La,$\dots$, Dy). The hexagonal \textsl{R}MnO$_3$ show both,
ferroelectric and magnetic order, but the respective ordering
temperatures differ by an order of magnitude, $T_{\rm FE}\simeq
1000$\,K and $T_{\rm N}\simeq 100$\,K.\cite{fiebig02a} In
contrast, \textsl{R}MnO$_3$ with \textsl{R} = Gd, Tb, and Dy
possess comparable transition temperatures for the magnetic and
the ferroelectric ordering.\cite{kimura03a,goto04a} Both,
TbMnO$_3$ and DyMnO$_3$ develop an incommensurate
antiferromagnetic order (ICAFM) below about 40~K. The
incommensurability continuously changes upon cooling and becomes
almost constant below $T_{\rm c}\simeq 28$~K and $\simeq 19$~K
for $R={\rm Tb}$ and Dy, respectively. The transition to this
long-wavelength incommensurate antiferromagnetic (LT-ICAFM) phase
is accompanied by ferroelectric (FE) ordering with a polarization
${\bf P}||c$, which flops to ${\bf P}||a$ above a critical
magnetic field applied along the $a$ or $b$ direction, while it
is suppressed for large fields along $c$.\cite{kimura05a}
GdMnO$_3$ also shows an ICAFM order below $T_{\rm N}\simeq 42$~K
and a second transition at $T_{\rm c}\simeq 23$~K. Based on the
observed weak ferromagnetism\cite{kimura03b,hemberger04b} and on
X-ray diffraction studies\cite{arima05a} a  canted A-type
antiferromagnetic ordering (cAFM) has been proposed for $T<T_{\rm
c}$, but a direct magnetic structure determination has not yet
been published. Both transition temperatures weakly increase in a
magnetic field.\cite{kimura05a} Concerning the FE polarization,
contradictory results have been reported. Kuwahara \textsl{et
al.}\cite{kuwahara05a} find a finite polarization below 13~K,
while Kimura \textsl{et al.}\cite{kimura05a} observe FE order
only between 5~K and 8~K. The magnitude of {\bf P} is much
smaller than in TbMnO$_3$ and DyMnO$_3$ and the direction is
${\bf P}||a$. Moreover, there are no magnetic-field-induced
polarization flops. Instead the polarization is stabilized by a
magnetic field along $b$, while it is immediately suppressed for
fields along $a$ and $c$.\cite{kimura05a,kuwahara05a}

In order to study the coupling of the various phase boundaries to
lattice degrees of freedom we have conducted high-resolution
measurements of thermal expansion and magnetostriction on
GdMnO$_3$. We find pronounced anomalies at all transitions, i.e.\
all transitions strongly couple to the lattice. Thus, our data
allow for a precise determination of the magnetic-field
temperature phase diagram and clearly reveal that the
ICAFM-to-cAFM as well as the FE transition are of first oder with
strong hysteresis. For low magnetic fields we find a down-bending
of the ICAFM-to-cAFM phase boundary and evidence for a
coexistence of both phases in this low-field range. This should
be taken into account for further experiments, in particular for
zero magnetic field. Probably, this may also explain the
contradictory results reported for
GdMnO$_3$.\cite{kimura05a,kuwahara05a,remarkoxygen}

%\section{Experimental Details}
The GdMnO$_3$ single crystal used in this study is a cuboid of
dimensions $1.7\times 2\times 1.45$\,mm$^3$ along the $a$, $b$,
and $c$ direction (\textsl{Pbnm} setting), respectively. It was
cut from a larger crystal grown by floating-zone melting. Phase
purity was checked by x-ray powder
diffraction.\cite{kadomtseva05a} Magnetization, resistivity and
specific heat data of the same crystal are reported in
Ref.~\onlinecite{hemberger04b}. The linear thermal expansion
$\alpha_i=\partial \ln L_i /
\partial T$ and magnetostriction $\Delta L_i(H)/L_i=[L_i(H)-L_i(0)] /
L_i(0) $ have been measured by a home-built high-resolution
capacitance dilatometer.\cite{lorenz97a} Here, $L_i$ denote the
lengths parallel to the different crystal axes $i=a$, $b$, and
$c$. In general, we studied the length changes in longitudinal
magnetic fields up to 14~T, i.e.\ $H||i$, and, in addition, we
measured $\alpha_b$ for $H||c$.

%\section{Results and Discussion}

\begin{figure}[tb]
\includegraphics[width=6.8cm,clip]{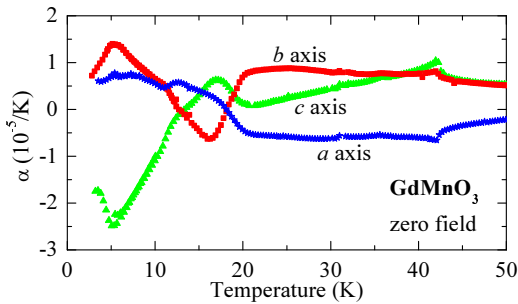}
\caption{(Color online) Thermal expansion of GdMnO$_3$ along $a$,
$b$, and $c$. \label{fig:zerofieldalpha}}
\end{figure}

Fig.~\ref{fig:zerofieldalpha} shows the zero-field $\alpha_i$ of
GdMnO$_3$ for $i=a$, $b$, and $c$, which are strongly anisotropic
and show several anomalies. The sharp anomalies around 41~K
signal the N\'{e}el transition of the Mn ions and their shape is
typical for a second-order phase transition. According to
previous publications further anomalies are expected at lower
$T$: (i) at the ICAFM-to-cAFM transition around $T_{\rm c}\simeq
23$~K, (ii) around $T_{\rm FE}\simeq 10$~K, where the FE ordering
sets in, and (iii) around 6~K due to the ordering of the Gd
moments. Indeed, there are pronounced anomalies around 6~K, small
ones around 10~K, and intermediate ones around 20~K with
different signs and magnitudes for the different $i$. However,
all these anomalies are rather broad making a clear
identification of transition temperatures difficult. This
drastically changes for finite magnetic fields.

\begin{figure}[tb]
\includegraphics[width=7.5cm,clip]{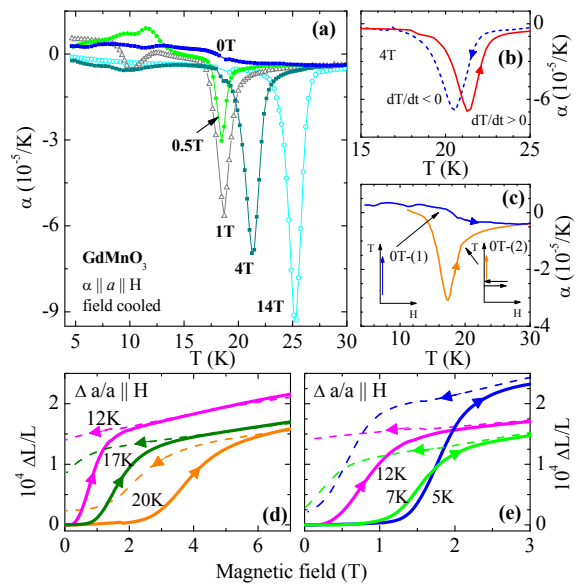}
\caption{(Color online) Panel (a): Field-cooled thermal expansion
$\alpha_a(T)$ for $H||a$ measured with increasing $T$. Panel~(b)
shows the hysteresis around $T_{\rm c}$ between the data obtained
with increasing and decreasing $T$.  Panel (c) compares two
zero-field $\alpha_a(T)$ curves for increasing $T$. The first
curve [0T-(1)] has been recorded after zero-field cooling the
sample and the second [0T-(2)] after cooling the sample to 12\,K
and subsequently applying and removing a magnetic field of 8~T.
The lower panels (d) and (e) present the zero-field cooled
magnetostriction $\Delta L_i(H)/L_i$ recorded with increasing
(solid lines) and decreasing (dashed)
field.\label{fig:aAxis_TADMS}}
\end{figure}

\begin{figure}[tb]
\includegraphics[width=7.cm,clip]{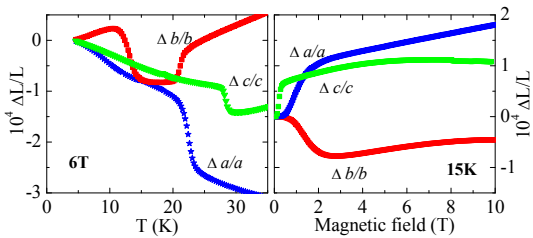}
\caption{(Color online) The relative length changes of the $a$,
$b$, and $c$ axes as a function of $T$ at constant field
($H=6$~T; left) and as a function of $H$ at constant temperature
($T= 15$~K; right). In all cases $H$ was applied parallel to the
measured direction. \label{fig:dLL_abc}}
\end{figure}

In Fig.~\ref{fig:aAxis_TADMS} we show $\alpha_a(T)$ for $H||a$.
With increasing magnetic field a broad anomaly shows up around
10~K, changes sign and smears out above 2~T. The origin of this
anomaly is unclear, it may be related to the complex interplay of
the magnetism of the Mn and Gd ions.\cite{hemberger04b} The most
drastic change occurs, however, around 18~K, where a huge anomaly
emerges between 0 and 1~T. This anomaly is close to the observed
ICAFM-to-cAFM transition at $T_{\rm c}$ and systematically shifts
to higher $T$ with further increasing field, in agreement with
the observed field dependence $T_{\rm
c}(H)$.\cite{kimura03a,kimura05a,arima05a} The negative sign of
the anomaly means that the transition from the cAFM to the ICAFM
phase, with increasing $T$, is accompanied by a pronounced
contraction of the $a$ axis (see also Fig.~\ref{fig:dLL_abc}).
The $\alpha_a(T)$ curves of Fig.~\ref{fig:aAxis_TADMS}a have been
recorded with increasing $T$ in the so-called field-cooled (FC)
mode, i.e.\ the field has been applied at $T\simeq 50$~K. We have
also recorded the data during the cooling runs.
Fig.~\ref{fig:aAxis_TADMS}b compares $\alpha_a(T)$ obtained with
increasing and decreasing $T$ for $H=4$~T. Obviously, there is a
strong hysteresis at the ICAFM-to-cAFM transition identifying
this transition as first-order type.

\begin{figure}[tb]
\includegraphics[width=7.2cm,clip]{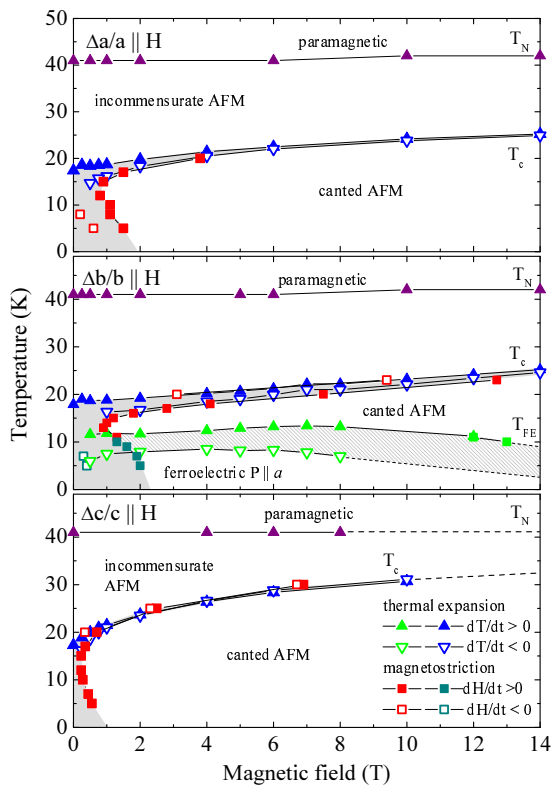}
\caption{(Color online) Phase diagrams of GdMnO$_3$ based on the
measurements of $\alpha_i$ (triangles) and $\Delta L_i(H)/L_i$
(squares) measured along $i=a$, $b$, and $c$ with $H||i$. Filled
and open symbols are obtained with increasing and decreasing $T$
or $H$, respectively. Hatched areas signal regions of strong
hysteresis, where different phases coexist. \label{fig:Phasdia}}
\end{figure}

The lower panels of Fig.~\ref{fig:aAxis_TADMS} present the
magnetostriction $\Delta L_a(H)/L_a $ for $H||a$ at constant $T$.
At 20\,K an anomalous expansion of the $a$ axis occurs around 4~T,
which is related to a transition from the ICAFM to the cAFM phase
as a function of increasing $H$. Both, the position of the
$\Delta L_a(H)/L_a$ anomaly in the $H-T$ plane and its magnitude
fit to the position and magnitude of the corresponding $\alpha_a
(T)$ anomaly due to the ICAFM-to-cAFM transition as a function of
decreasing $T$. The $\Delta L_a(H)/L_a$ curves obtained with
increasing and decreasing $H$ also show a hysteresis at the
ICAFM-to-cAFM transition. With decreasing $T$ the anomaly of
$\Delta L_a(H)/L_a $ shifts towards lower field. Around 12~K an
anomalous expansion of the $a$ axis is observed with increasing
$H$, however, this expansion is not reversed upon decreasing the
field. Due to the large hysteresis GdMnO$_3$ remains in the cAFM
phase after the magnetic field is switched off. With further
decreasing $T$, the anomaly of $\Delta L_a(H)/L_a $ again shifts
to higher field and at 5~K the anomalous expansion occurring with
increasing $H$ is reversed again with decreasing field.
Apparently, we have traced the field-induced transition to the
cAFM phase down to our lowest $T$. This means that the
ICAFM-to-cAFM phase boundary $T_{\rm c}(H)$ shows a clear
down-bending in the low-field range (see Fig.~\ref{fig:Phasdia}).
In order to verify this, we have carried out the following
zero-field $\alpha_a(T)$ measurement: after cooling the sample to
12~K in zero field we have applied a magnetic field of 8\,T in
order to enter the cAFM phase and due to the hysteresis of the
ICAFM-to-cAFM transition the sample should remain in the cAFM
phase after removing the field again. Fig.~\ref{fig:aAxis_TADMS}c
shows a comparison of this zero-field $\alpha_a(T)$ [labeled as
0T-(2)] with the conventional FC $\alpha_a(T)$ [0T-(1)]. In
contrast to the 0T-(1) curve, the 0T-(2) measurement shows a
sharp peak around $T_{\rm c}$, which is comparable to the anomaly
of the finite-field $\alpha_a(T)$ curves. This clearly confirms
the down bending of the ICAFM-to-cAFM phase boundary. Thus, the
(pure) cAFM phase of GdMnO$_3$ cannot be reached by cooling the
crystal in zero field. The broad anomalies in the zero-field
$\alpha_i(T)$ curves (see Fig.~\ref{fig:zerofieldalpha}) probably
arise from a partial ICAFM-to-cAFM transition and suggest that
both phases coexist in the low-field region.

In Fig.~\ref{fig:dLL_abc} we compare the relative length changes
as a function of field and temperature for all three axes. The
$c$ axis behaves similar to the $a$ axis. The anomalies of $c$
have the same signs but smaller magnitudes as compared to those
of $a$ and, moreover, their position for the same field
(temperature) is located at higher $T$ (lower $H$). Since in all
cases we applied $H|| L_i$, the latter difference signals the
anisotropy with respect to the field direction. Our results (not
shown) for $\alpha_c (T)$ and $\Delta L_c(H)/L_c$ for other
fields and temperatures, respectively, are very similar to those
obtained for the $a$ axis. In particular, the distinct expansion
at the ICAFM-to-cAFM transition as a function of field or
temperature is only present for $H\gtrsim 0.2$~T. The hysteresis
is also strong for $H||c$, but less pronounced than for $H||a$
(see Fig.~\ref{fig:Phasdia}). A different phenomenology is
observed for $\Delta L_b/L_b$ for $H||b$. Firstly, the anomalies
at the ICAFM-to-cAFM transition are of opposite signs and,
secondly, an additional anomaly occurs at a lower temperature.

\begin{figure}[tb]
\includegraphics[width=7.5cm,clip]{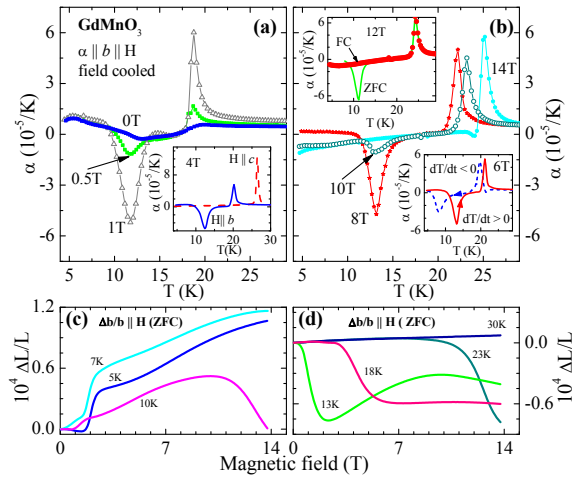}
\caption{(Color online) Field-cooled measurements of $\alpha_b(T)$
in fields up to 1\,T (a) and between 8\,T and 14\,T (b). The inset
of panel~(a) compares $\alpha_b(T)$ for $H=4$~T applied along $b$
(solid line) and $c$ (dashed). The upper inset of panel~(b) shows
$\alpha_b(T)$ for $H=12$~T ($H||b$; $dT/dt>0$) obtained in a FC
(symbols) and a ZFC run (line). The lower inset displays the
hysteresis between $\alpha_b(T)$ for $H=6$~T recorded with
$dT/dt>0$ (solid line) and $dT/dt<0$ (dashed). Panels~(c) and (d)
show the ZFC magnetostriction obtained for $dH/dt >0$.
\label{fig:bAxis_TAD}}
\end{figure}

In Fig.~\ref{fig:bAxis_TAD} we present $\alpha_b(T)$ for
different $H||b$. Again, a very pronounced anomaly evolves around
$T_{\rm c}\simeq 18$~K in small fields and shifts to higher $T$
with further increasing field. The behavior of $\alpha_b $ is
analogous to our results on $\alpha_a$ and $\alpha_c$, only the
signs of the anomalies are different. In contrast to $\alpha_a$
and $\alpha_c$, however, an additional, very pronounced anomaly
develops around 12~K in finite fields. Based on the polarization
data of Ref.~\onlinecite{kimura05a}, we attribute this anomaly to
the FE ordering at $T_{\rm FE}$. As a further verification we
have also measured $\alpha_b$ in a transverse magnetic field
$H||c$ and did not find such an additional anomaly (see inset of
Fig.~\ref{fig:bAxis_TAD}a). As shown in the lower inset of
Fig.~\ref{fig:bAxis_TAD}b the transition to the FE phase is also
of first-order type with a broad hysteresis. With further
increase of the magnetic field above 8\,T the magnitude of the
$T_{\rm FE}$ anomaly decreases again and vanishes around 12~T.
This disappearance is a consequence of the strong hysteresis of
the FE transition and occurs only in the FC measurements of
$\alpha_b$. The upper inset of Fig.~\ref{fig:bAxis_TAD}b compares
the FC and zero-field-cooled (ZFC) measurements of $\alpha_b $
for $H = 12$~T~$||\,b$. The ZFC curve displays a large anomaly at
$T_{\rm FE}$, which is absent in the FC curve. This difference
arises from a decreasing $T_{\rm FE}(H)$ with increasing field.
When the lower $T_{\rm FE}(H)$ is smaller than our lowest
measurement temperature of $\simeq 4.5$~K the FE phase cannot be
reached in a FC run and, consequently, there is no anomaly in the
subsequent $\alpha_b(T)$ measured with increasing $T$. The FE
phase can, however, be entered by cooling the sample in a lower
field, e.g.\ 6~T (see lower inset of Fig.~\ref{fig:bAxis_TAD}b),
or in a ZFC run when the field is applied at the lowest
temperature. Due to the large hysteresis the sample remains in
the FE phase up to high fields and therefore the ZFC
$\alpha_b(T)$ curves show anomalies at the upper $T_{\rm FE}(H)$
when the FE phase is left upon heating.

The lower panels (c) and (d) of Fig.~\ref{fig:bAxis_TAD} show the
magnetostriction. In agreement with the negative signs of the
$\alpha_b$ anomalies we observe a pronounced contraction at the
field-induced ICAFM-to-cAFM transition for  $T \gtrsim 13$~K. At
lower $T$ we still find clear anomalies, however, of opposite
sign. This sign change results from the fact that below about
10~K the FE phase is entered, which has a significantly longer
$b$ axis than both, the cAFM phase and a low-$T$ extrapolation of
the ICAFM phase (see Fig.~\ref{fig:dLL_abc}). The strong decrease
of $\Delta b/b$ at 10~K for $H\gtrsim 10$~T signals a
field-induced FE-to-cAFM transition due to the above-mentioned
negative slope of the $T_{\rm FE}(H)$ boundary.

Based on our measurements of $\alpha_i$ and $\Delta L_i(H)/L_i$
with $H||i$ for all three crystallographic axes $i$ we derive the
phase diagrams presented in Fig.~\ref{fig:Phasdia}. The weak field
dependence of $T_{\rm N} \simeq 42$~K well agrees with previous
results.\cite{kimura05a} The new feature of our phase diagram is
the down-bending of the ICAFM-to-cAFM phase boundary at low
magnetic field. This transition exhibits a strong hysteresis and
we conclude that the ICAFM and the cAFM phases coexist in the
low-field low-temperature region. Such a coexistence can
naturally explain why the zero-field $\alpha_i(T)$ curves only
show some broad anomalous features around 18~K instead of the
distinct anomalies which signal the transition from the cAFM to
the ICAFM phase for larger fields. This conclusion is also
supported by measurements of the polarization showing that the
magnitude of $\bf P$ in this hysteretic region is much smaller
than both, $\bf P$ for larger $H||b$ and $\bf P$ of $R$MnO$_3$
with $R={\rm Tb}$ and Dy.\cite{kimura05a}

As shown in Fig.~\ref{fig:bAxis_TAD}a the magnitudes of the
$\alpha_b$ anomalies at $T_{\rm FE}$ and $T_{\rm c}$
simultaneously evolve between 0 and 1~T. This correlation
suggests that the ICAFM-to-cAFM transition is a precondition for
the FE ordering. As mentioned above the notation 'cAFM' should be
treated with some caution, because the magnetic structure of
GdMnO$_3$ has not yet been unambiguously determined. In a
simplified picture, the proposed structure can be described as
follows: the Mn moments are oriented approximately along $b$ with
some canting towards $c$, along the $a$ ($c$) direction
neighboring moments are essentially parallel (antiparallel) with
respect to each other, and along $b$ an incommensurate modulation
of the moments is present above and vanishes below $T_{\rm c}$.
This view is supported by the phase diagram, since the
stabilization of the cAFM phase is most pronounced for $H||c$ as
it is expected for a cAFM phase with a weak ferromagnetic moment
pointing along $c$ already in zero field. A field along $a$ also
points approximately perpendicular to the Mn moments, and one can
therefore expect that the Mn moments (and also the weak
ferromagnetic moment) are slightly canted towards $a$. However, a
more drastic change is expected for $H||b$, since 50\% of the Mn
moments are oriented roughly antiparallel to $H$. In simple
antiferromagnets this configuration usually leads to a spin-flop
transition, i.e.\ the Mn moment would jump to the $ac$ plane and
cant towards $b$. Such a configuration would be analogous to
those for $H||a$ or $c$ and according to
Ref.~\onlinecite{mostovoy05a} none of these configurations would
lead to a finite polarization via a coupling of the magnetic and
ferroelectric order parameters. However, in view of the more
complex magnetic structures observed in neighboring $R$MnO$_3$,
one may speculate that a similar complex structure could be
induced in GdMnO$_3$ for $H||b$, and this might explain the finite
FE polarization via a coupling between AFM and FE order
parameters.\cite{mostovoy05a,kenzelmann05a} Hence, there is need
for a detailed determination of the magnetic structure of
GdMnO$_3$.

In summary, we have presented a study of the magnetic-field
temperature phase diagram of GdMnO$_3$ via thermal expansion and
magnetostriction measurements. We find that both, the
ICAFM-to-cAFM as well as the FE transition are of first-order
type and strongly hysteretic. The hysteresis is most pronounced
in the low-field range. We find a down-bending of the
ICAFM-to-cAFM phase boundary and evidence for coexisting ICAFM
and cAFM phases in this low-field range.

We acknowledge fruitful discussions with D.\,Khomskii. This work
was supported by the Deutsche Forschungsgemeinschaft through
SFB~608. JAM is supported by the Alexander von Humboldt Stiftung.

%\bibliography{p:/Preload,RMnO3,remarks}

\begin{thebibliography}{14}
\bibitem[{rem()}]{remarkoxygen}
\bibinfo{howpublished}{According to Ref.~\onlinecite{goto05a} this discrepancy
  could also arise from weak oxygen off-stoichiometries in different crystals.}
\end{thebibliography}

\end{document}